\documentstyle[aps,prl,preprint,floats,epsfig]{revtex}

\begin{document}

\preprint{\tighten\vbox{\hbox{\hfil CLNS 01-1732}
			\hbox{\hfil CLEO 01-08}}}

\title{\bf {Experimental Investigation of the Two-photon Widths of the
$\chi_{c0}$ and the $\chi_{c2}$ Mesons}}


\author{CLEO Collaboration}
\date{\today}

\maketitle
\tighten

\begin{abstract}

Using 12.7 ${\rm fb}^{-1}$ of data collected with the CLEO detector at CESR, we observed two-photon production of the $c \bar c$ states $\chi_{c0}$ and $\chi_{c2}$ in their decay to $\pi^+ \pi^- \pi^+ \pi^-$. 
We measured 
$\Gamma_{\gamma\gamma}(\chi_{c}) \times {\mathcal B}(\chi_{c} \rightarrow 
\pi^+\pi^-\pi^+\pi^-)$ to be 75 $\pm$ 13 (stat) $\pm$ 8 (syst) eV 
for the $\chi_{c0}$ and 6.4 $\pm$ 1.8 (stat) $\pm$ 0.8 (syst) eV 
for the $\chi_{c2}$, implying
$\Gamma_{\gamma\gamma}(\chi_{c0}) = 3.76 \pm 
0.65 ({\rm stat}) \pm 0.41 ({\rm syst}) \pm 1.69 ({\rm br})$ keV and 
$\Gamma_{\gamma\gamma}(\chi_{c2}) = 0.53 \pm 0.15 ({\rm stat}) \pm 
0.06 ({\rm syst}) \pm 0.22 ({\rm br})$ keV. 
Also, cancelation of dominant experimental and theoretical uncertainties permits a precise comparison of $\Gamma_{\gamma\gamma}(\chi_{c0})/ \Gamma_{\gamma\gamma}(\chi_{c2})$, evaluated to be $7.4 \pm 2.4 ({\rm stat}) \pm 0.5 ({\rm syst}) \pm 0.9 ({\rm br)}$, with QCD-based predictions.
\end{abstract}
\newpage

{
\renewcommand{\thefootnote}{\fnsymbol{footnote}}

\begin{center}
B.~I.~Eisenstein,$^{1}$ J.~Ernst,$^{1}$ G.~E.~Gladding,$^{1}$
G.~D.~Gollin,$^{1}$ R.~M.~Hans,$^{1}$ E.~Johnson,$^{1}$
I.~Karliner,$^{1}$ M.~A.~Marsh,$^{1}$ C.~Plager,$^{1}$
C.~Sedlack,$^{1}$
% ...
\end{center}
 
\small
\begin{center}
$^{1}${University of Illinois, Urbana-Champaign, Illinois 61801}\\
% ...
\end{center}

\setcounter{footnote}{0}
}
\newpage


In this Letter, we report a study of two-photon production of the C-even $1^3 P$ charmonium states $\chi_{c0}$ and $\chi_{c2}$ using the CLEO detector at the Cornell Electron Storage Ring (CESR). 
The charmonium system is an ideal testing ground for quantum chromodynamics (QCD). 
Perturbative QCD (PQCD) provides predictions for the two-photon widths 
$\Gamma_{\gamma\gamma}(\chi_{c0})$ and $\Gamma_{\gamma\gamma}(\chi_{c2})$.
These predictions involve charmed quark mass factors, 
non-perturbative factors, and wave-function dependence, 
all of which cancel in the ratio 
$\Gamma_{\gamma\gamma}(\chi_{c0}) / \Gamma_{\gamma\gamma}(\chi_{c2})$.
The experimentally measured quantities are the two products
$\Gamma_{\gamma\gamma}(\chi_{c}) \times {\mathcal B}(\chi_{c} \rightarrow 
\pi^+\pi^-\pi^+\pi^-)$, $\chi_c \equiv \chi_{c0}$ or $\chi_{c2}$.
Systematic uncertainties in the measurements mostly cancel in the ratio
of these products.
Further, the contribution to the uncertainty on
$\Gamma_{\gamma\gamma}(\chi_{c0}) / \Gamma_{\gamma\gamma}(\chi_{c2})$
from the uncertainty on the ratio
${\mathcal B}(\chi_{c0} \rightarrow \pi^+\pi^-\pi^+\pi^-) /
{\mathcal B}(\chi_{c2} \rightarrow \pi^+\pi^-\pi^+\pi^-)$ is greatly
reduced because both branching fractions have been measured
in the same experiment leading to cancelation of 
their systematic uncertainties in the ratio.
Thus the ratio
$\Gamma_{\gamma\gamma}(\chi_{c0}) / \Gamma_{\gamma\gamma}(\chi_{c2})$
affords a more precise comparison of theory and experiment than do the
individual $\Gamma_{\gamma\gamma}(\chi_{c})$.

The two-photon width of a $\chi_c$ meson can be determined from a measurement of its two-photon cross section.
The ratio of the two-photon width to the two-gluon width of a $\chi_c$ meson can be calculated in PQCD with reduced uncertainties due to cancelation of charmed quark mass factors, non-perturbative factors, and wave function dependence. 
In next-to-leading order (NLO) PQCD one obtains the following relationships \cite{Kwong} :
\begin{equation}{\frac{\Gamma_{\gamma\gamma}(\chi_{c0})}
{\Gamma_{gg}(\chi_{c0})}} = 
\frac{8 \alpha^2}{9 \alpha_s^2} \frac{(1+0.18\alpha_s/\pi)}
{(1 + 9.5 \alpha_s/\pi)},\end{equation}
\begin{equation}{\frac{\Gamma_{\gamma\gamma}(\chi_{c2})}
{\Gamma_{gg}(\chi_{c2})}} = 
\frac{8 \alpha^2}{9 \alpha_s^2} \frac{(1-5.3\alpha_s/\pi)}
{(1 - 2.2 \alpha_s/\pi)}.\end{equation}

The width of the $\chi_{c0}$ meson can be assumed to be dominated by its two-gluon component, so $\Gamma_{gg}(\chi_{c0}) \approx \Gamma_{\rm tot}(\chi_{c0}) = 14.9^{+2.6}_{-2.3}$ MeV \cite{PDG}. 
Using a value of the strong coupling constant $\alpha_s = 0.28$ \cite{Kwong}, one obtains the NLO PQCD prediction $\Gamma_{\gamma\gamma}(\chi_{c0}) = 5.0 \pm 0.8$ keV. 
Due to the uncertainty in the charm mass scale, we also calculate the NLO PQCD prediction at $\alpha_s = 0.35$ and find $\Gamma_{\gamma\gamma}(\chi_{c0}) = 2.9 \pm 0.5$ keV.
A measurement reported in a thesis gave $\Gamma_{\gamma\gamma}(\chi_{c0}) = 4.0 \pm 2.8$ keV \cite{Lee}. The E835 collaboration reported an upper limit of $\Gamma_{\gamma\gamma}(\chi_{c0}) \le 3.47$ keV (95\% C.L.) \cite{E835}.
	
The two-gluon component of the $\chi_{c2}$ width can be extracted from its width $\Gamma_{\rm tot}(\chi_{c2}) = 2.00 \pm 0.18$ MeV \cite{PDG} by subtracting the radiative width $\Gamma (\chi_{c2} \rightarrow \gamma J/\psi)$ and the color-octet width in which the $\chi_{c2}$ decays via three gluons to light hadrons. This latter contribution has been shown in Ref. \cite{Mangano} to be equal to the hadronic width of the $\chi_{c1}$ meson. 
In the case of the $\chi_{c2}$ meson, the color octet contribution to the full width is significant; for the $\chi_{c0}$ meson, it is negligible.
Hence, one obtains the NLO PQCD prediction 
$\Gamma_{\gamma\gamma}(\chi_{c2}) = 0.47 \pm 0.04$ keV 
using a value of $\alpha_s = 0.28$. 
This prediction becomes $\Gamma_{\gamma\gamma}(\chi_{c2}) = 0.25 \pm 0.02$ keV for $\alpha_s = 0.35$. 
The Particle Data Group (PDG) value of $\Gamma_{\gamma\gamma}(\chi_{c2})$ 
is 0.47 $\pm$ 0.17 keV \cite{PDG}, 
where the uncertainty includes a scale factor of 1.9 to account for the
poor consistency among the various measurements. The recent measurement of $\Gamma_{\gamma\gamma}(\chi_{c2}) = 0.270 \pm 0.049 ({\rm stat}) \pm 
0.033 ({\rm syst})$ keV \cite{E835} by the E835 collaboration was not included in this average.

The comparison of the experimental results for $\Gamma_{\gamma\gamma}(\chi_{c0})$ and $\Gamma_{\gamma\gamma}(\chi_{c2})$ with the theoretical predictions is hampered by the lack of precision in the measured $\chi_c$ hadronic branching fractions, which are needed to extract the two-photon widths. 
Exploiting the fact that the $\chi_{c0}$ and the $\chi_{c2}$ mesons were both detected in their decay into $\pi^+\pi^-\pi^+\pi^-$, we calculate the ratio of their two-photon widths where the dominant branching fraction uncertainties cancel. 
In addition, some of the dominant systematic uncertainties in the $\Gamma_{\gamma\gamma}(\chi_{c0})$ and $\Gamma_{\gamma\gamma}(\chi_{c2})$ measurements cancel in the ratio because the $\chi_{c0}$ and the $\chi_{c2}$ mesons were detected in the same experiment. 
The NLO PQCD prediction for the ratio of the two-photon widths is \cite{Kwong}
\begin{equation}{\frac{\Gamma_{\gamma\gamma}(\chi_{c0})}
{\Gamma_{\gamma\gamma}(\chi_{c2})}} = 
\frac{15}{4} \frac{(1 + 0.18 \alpha_s/\pi)}{(1-5.3\alpha_s/\pi)},\end{equation}
Here too, charmed quark mass factors have canceled, and we have assumed $|\Psi'_{\chi_{c0}}(0)|^2 = |\Psi'_{\chi_{c2}}(0)|^2$  in the non-relativistic limit, where $\Psi'_{\chi_{c}}(0)$ denotes the derivative of the $\chi_c$ wave function at the origin. 
Relativistic corrections due to explicit modification of the meson decay amplitudes decrease the above ratio whereas corrections arising from the modification of the $\chi_c $ wave-functions increase the above ratio \cite{Huang}.
In NLO PQCD, predictions for this ratio range from 7 to 9 when $\alpha_s$ varies from 0.28 to 0.35. 
Experimentally, this ratio is found to be 8.7 $\pm$ 6.9, using the single available $\Gamma_{\gamma\gamma}(\chi_{c0})$ measurement \cite{Lee} and the PDG value of $\Gamma_{\gamma\gamma}(\chi_{c2})$ \cite{PDG}. 
Thus a precise measurement of the individual two-photon widths of the $\chi_{c0}$ and the $\chi_{c2}$ mesons and, more importantly, their ratio are of interest for a comparison with theory. 

At CESR, which operates at and near the $\Upsilon(4S)$ resonance, C-even charmonium states are produced through the fusion of two space-like photons, radiated by the 5.3 GeV $e^+$ and $e^-$ beams. 
The data used in this study correspond to an integrated luminosity of 
12.7 ${\rm fb}^{-1}$ and were collected with two configurations (CLEO II \cite{CLEOII} and CLEO II.V \cite{CLEOII.V}) of the CLEO detector.
Approximately one third of the data was taken with the CLEO II configuration.  The detector components most useful for this study were the concentric tracking devices for charged particles, operating in a 1.5 T 
magnetic field generated by a superconducting solenoid. 
For CLEO II, this tracking system consisted of a 6-layer straw tube chamber, 
a 10-layer precision drift chamber, and a 51-layer main drift chamber. 
The main drift chamber also provided measurements 
of the specific ionization loss, $dE/dx$, used for particle identification. 
For CLEO II.V, the straw tube chamber was replaced by a 3-layer, 
double-sided silicon vertex detector and the gas in the main drift chamber 
was changed from a 50:50 mixture of argon-ethane to a 60:40 helium-propane 
mixture. 
These changes gave rise to a significant improvement in the momentum 
and $dE/dx$ resolutions for charged tracks. 
Photons were detected using the high-resolution electromagnetic 
calorimeter consisting of 7800 CsI crystals. 
The Monte Carlo simulation of the CLEO detector response 
was based upon GEANT \cite{GEANT}. 
Simulated events were processed in the same manner as the data to determine the $\chi_{c} \rightarrow \pi^+ \pi^- \pi^+ \pi^-$ detection efficiencies and the $\pi^+ \pi^- \pi^+ \pi^-$ mass resolutions at the two $\chi_{c}$ meson masses.

In the two-photon process $e^+e^-$ $\rightarrow$ $e^+e^-\gamma\gamma$ $\rightarrow$ $e^+e^-\chi_c$ $\rightarrow$ $e^+e^-\pi^+\pi^-\pi^+\pi^-$, the photon propagators dictate that the photons are almost real 
(``on shell''). 
Therefore the incident leptons are scattered at very small angles with the beam and remain undetected. 
Such ``untagged'' events typically 
have low transverse momentum ($p_T \equiv |\sum_{i}\ \vec{p}_{T_i}|$, $i = 1-4$)
and low visible energy. 

The events were required to have exactly four charged tracks with zero total charge. 
The background from processes other than two-photon production was suppressed by requiring that the $\chi_c$ candidate reconstructed from these four charged tracks has transverse momentum less than 0.4 GeV/c and that the visible energy in the event be less than 6.0 GeV. 
A $\chi^2$ probability was constructed using the $dE/dx$ information from the four tracks, and required to be greater than 10\%.
Also, because the final state had no expected energy deposits in the calorimeter from neutral particles, the total calorimeter energy in the event not matched to charged tracks was required to be less than 0.6 GeV. 
We required that at least two of the four tracks traversed all layers 
of the tracking volume to achieve a well-modeled trigger simulation
in the Monte Carlo.

\begin{figure}[h] 
  \centerline{\epsfig{file=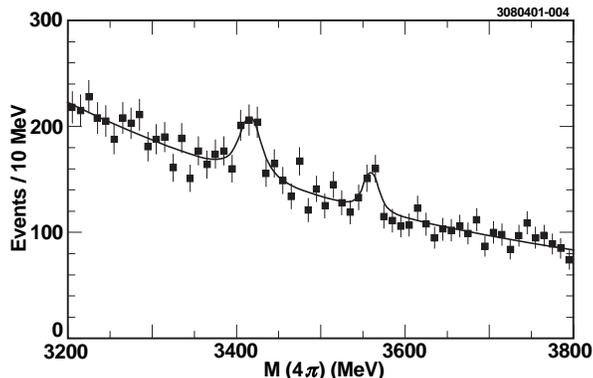, height=2.0in}}
  \caption{The $\pi^+\pi^-\pi^+\pi^-$ invariant mass (data point with
errors). The solid line is the fit with a $\chi^2$/d.o.f. = 44/54.}
  \label{mass}
\end{figure}

The invariant mass distribution of the selected four-pion events is shown in Figure \ref{mass}. 
Enhancements near the known $\chi_{c0}$ and $\chi_{c2}$ masses are seen upon a smooth background. 
To obtain the number of signal events and mass values, 
the mass distribution was fitted as follows. 
The background was fitted with a power law function: 
$A \cdot W_{\gamma\gamma}^n$, with $W_{\gamma\gamma}$ 
the $\pi^+\pi^-\pi^+\pi^-$ invariant mass and $A$ and $n$ 
parameters to be fitted.
The $\chi_{c0}$ signal was fitted with a spin-0 relativistic 
Breit-Wigner function with a width equal to the PDG value of 14.9 MeV
(describing the natural line shape) 
convolved with a double Gaussian function 
(describing the detector resolution). 
The $\chi_{c2}$ signal was fitted using only a double Gaussian resolution
because the PDG value of 2.0 MeV for its width is negligible in comparison 
to the mass resolution.
The parameters of the resolution functions were determined from 
Monte Carlo samples of the $\chi_{c0}$ and $\chi_{c2}$, 
generated with zero width. 
The invariant mass resolution was approximately 9 MeV for both the 
resonances. 
We performed a simultaneous, binned, maximum-likelihood fit 
to the invariant mass distribution using these three functions. 

From the fit, we obtained 234 $\pm$ 40 candidate $\chi_{c0}$ 
events and 89 $\pm$ 25 candidate $\chi_{c2}$ events, with masses of 3416.5 $\pm$ 3.0 MeV and 3559.9 $\pm$ 2.9 MeV, respectively. 
Our measured masses are consistent with the PDG values, 3415.0 $\pm$ 0.8 MeV and 3556.18 $\pm$ 0.13 MeV \cite{PDG} respectively.

Comparison of the two-photon crosssection of a resonance measured 
from data with that estimated from Monte Carlo for a given two-photon width allows extraction of the $\Gamma_{\gamma\gamma}(\chi_c)$. 
The Monte Carlo crosssection ($\sigma_{\rm MC}$) estimation was based upon the 
BGMS formalism \cite{BGMS}. 
The cross-section in data was calculated by dividing 
the number of events from the fit by the detection efficiency,
the integrated luminosity, and the branching fraction 
${\mathcal B} (\chi_c \rightarrow \pi^+\pi^-\pi^+\pi^-)$.
The detection efficiencies were 19.6\% and 21.7\% for the 
$\chi_{c0}$ and the $\chi_{c2}$, respectively.
Due to the large uncertainties in the branching fractions, 
we preferred to express our primary results as the products of the 
branching fraction and the two-photon width. 
We obtained 
$\Gamma_{\gamma\gamma}(\chi_{c0}) \times 
{\mathcal B}(\chi_{c0} \rightarrow \pi^+\pi^-\pi^+\pi^-) = 
75 \pm 13 ({\rm stat}) \pm 8 ({\rm syst}) $ eV and 
$\Gamma_{\gamma\gamma}(\chi_{c2}) \times 
{\mathcal B}(\chi_{c2} \rightarrow \pi^+\pi^-\pi^+\pi^-) = 
6.4 \pm 1.8 ({\rm stat}) \pm 0.8 ({\rm syst})$ eV. 
From these primary results we obtained the ratio 
$\Gamma_{\gamma\gamma}(\chi_{c0})/\Gamma_{\gamma\gamma}(\chi_{c2}) = 
7.4 \pm 2.4 ({\rm stat}) \pm 0.5 ({\rm syst}) \pm 0.9 ({\rm br})$, 
where the last uncertainty corresponds to branching fraction 
uncertainties that do not cancel in the ratio. 
Using the known branching fractions, ${\mathcal B} (\chi_{c0} \rightarrow \pi^+\pi^-\pi^+\pi^-) = 2.0 \pm 0.9$ \% and ${\mathcal B} (\chi_{c2} \rightarrow \pi^+\pi^-\pi^+\pi^-) = 1.2 \pm 0.5$ \% \cite{PDG}, 
we obtained the two-photon widths as 
$\Gamma_{\gamma\gamma}(\chi_{c0}) = 
3.76 \pm 0.65 ({\rm stat}) \pm 0.41 ({\rm syst}) \pm 1.69 ({\rm br})$ keV and 
$\Gamma_{\gamma\gamma}(\chi_{c2}) = 
0.53 \pm 0.15 ({\rm stat}) \pm 0.06 ({\rm syst}) \pm 0.22 ({\rm br})$ keV, 
where the last uncertainties correspond to the systematic uncertainties 
arising from the hadronic branching fraction.

Sources of systematic uncertainty in the measured product branching fractions 
are summarized in Table I.
The uncertainties which were correlated are marked by an asterisk.
The uncertainties were added in quadrature to obtain the total 
systematic uncertainty. 
The systematic uncertainty 
of the ratio of the two-photon widths takes into account
correlations among the uncertainties. 
Systematic uncertainties in this ratio were separately 
estimated for our experimental measurement and the branching fraction 
measurements from other experiments \cite{MRK1,BES,XBAL}. 
Though the individual measurements of  
${\mathcal B}(\chi_{c0} \rightarrow \pi^+\pi^-\pi^+\pi^-)$ and the 
${\mathcal B}(\chi_{c2} \rightarrow \pi^+\pi^-\pi^+\pi^-)$ 
have poor consistency, the ratios of these two branching fractions are consistent.

\begin{center}
\begin{table}
\caption{Systematic uncertainties in the two 
$\Gamma_{\gamma\gamma}(\chi_c) \times {\mathcal B}(\chi_c \rightarrow \pi^+\pi^-\pi^+\pi^-)$ measurements.  
The total systematic uncertainties were obtained by adding the individual 
contributions in quadrature.
Most systematic uncertainties are pairwise correlated and are marked with an asterisk.}
\smallskip
\begin{tabular}{l c c}
Source of Uncertainty           & $\chi_{c0}$ (\%) & $\chi_{c2}$ (\%)\\
\hline
Event $p_T$*      		&  7  & 7  \\
Particle Identification*	&  4  & 6  \\
Unmatched Neutral Energy*       &  4  & 4  \\
Helicity			&  0  & 4  \\
Mass Calibration*		&  1  & 3  \\
Width of Resonance		&  4  & 0  \\
$\rho^0$ Substructure*		&  2  & 2  \\
Trigger*			&  2  & 2  \\
Tracking*                       &  2  & 2  \\
Detector Resolution*		&  1  & 1  \\
Signal Shape Parameter*		&  2  & 2  \\
$\sigma_{\rm MC}$ (inc pole mass)*	&  1  & 1  \\
Luminosity*			&  1  & 1  \\
\hline
Total                           &  11 & 12\\
\end{tabular}
\end{table}
\end{center}

The systematic uncertainties were dominated by the uncertainties in the simulation of the signal $p_T$ distribution, the particle identification procedure and the simulation of the unmatched neutral energy. 

In our cross-section estimate from Monte Carlo we had assumed that the $\chi_{c2}$ meson was produced only in the helicity 2 state. 
The helicity 0 state has zero two-photon width in the non-relativistic approximation, while in relativistic models it is predicted to be up to 4\% of the helicity 2 state \cite{Huang}. 
We allowed the mass of each resonance to vary between our measurement 
and the PDG value to estimate the systematic uncertainty from uncertainty in the mass scale. We similarly varied the width of the $\chi_{c0}$ within its known uncertainty.
Using the known branching fractions of the $\chi_{c0}$ and $\chi_{c2}$ \cite{MRK1} into $\rho^0 \pi^+ \pi^-$ and estimating the difference in detection efficiencies using Monte Carlo we assigned the systematic uncertainty due to the presence of resonant $\rho^0$ substructure in the $\chi_c$ decay. 
Ref. \cite{MRK1} had reported no significant $\rho^0 \rho^0$ sub-structure in the $\chi_{c0}$ and $\chi_{c2}$ decay.

We investigated the possible effects of interference between the four-pion decay of the $\chi_c$ states and non-resonant production of four pions in two-photon interaction.
From Monte Carlo simulation we found that a fifth of the selected events were from annihilation $\tau$-pair production. 
Another one-third were estimated to be not of two-photon origin based on the candidate transverse momentum distribution. 
Neither of these two classes of events can interfere with signal events. 
Interference can distort the signal shape derived from the signal Monte Carlo. 
We observed no such distortion demonstrated by the good fit $\chi^2$ in the two regions of the resonances: 12.4 for 12 degrees of freedom.

The systematic uncertainty of 0.5 in the ratio $\Gamma_{\gamma\gamma}(\chi_{c0}) / \Gamma_{\gamma\gamma}(\chi_{c2})$ takes the correlations marked by an asterisk in Table I into account. The systematic uncertainty of 0.9 from the ratio of branching fractions was calculated taking into account correlations as ascertained from Ref. \cite{MRK1} and \cite{BES}.
The ratio of the branching fractions has significantly lower relative uncertainty than the branching fractions themselves due to partial cancelation of correlated systematic uncertainties. 
The branching fractions ${\mathcal B} (\psi(2S) \to \gamma \chi_c)$ needed
were taken from Ref. \cite{XBAL}, taking into account correlations between them.

\begin{figure}[h] 
  \centerline{\epsfig{file=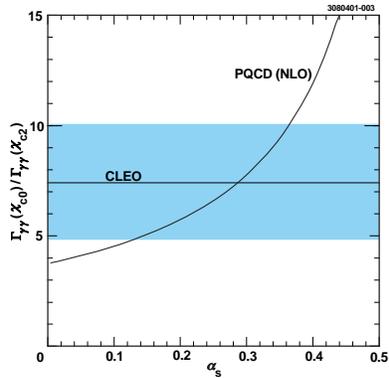, height=2.0in}}
  \caption{Comparison of the measurement of the ratio
$\Gamma_{\gamma\gamma}(\chi_{c0}) / \Gamma_{\gamma\gamma}(\chi_{c2})$ 
(shaded region corresponds to $\pm$ 1 $\sigma$) with its prediction from NLO PQCD as function of $\alpha_s$.}
  \label{ratio}
\end{figure}

Our result $\Gamma_{\gamma\gamma}(\chi_{c0}) / \Gamma_{\gamma\gamma}(\chi_{c2})
= 7.4 \pm 2.4 ({\rm stat}) \pm 0.5 ({\rm syst}) \pm 0.9 ({\rm br})$ represents a significant improvement upon the current value of 8.7 $\pm$ 6.9 and is compared with
the NLO PQCD prediction (Eq. 3) in Figure \ref{ratio}.
There is good agreement for reasonable values of $\alpha_s$; however one wishes for increased theoretical and experimental precision.
Our $\Gamma_{\gamma\gamma}(\chi_{c0})$ measurement of 3.76 $\pm$ 0.65 (stat) $\pm$ 0.41 (syst) $\pm$ 1.69 (br) keV is a significant improvement in experimental precision over the single available measurement of 4.0 $\pm$ 2.8 keV \cite{Lee}, but limited by poor knowledge of the branching fraction. 
The NLO PQCD prediction of 3 - 5 keV is consistent with our measurement. 
Our $\Gamma_{\gamma\gamma}(\chi_{c2})$ measurement of 0.53 $\pm$ 0.15 (stat) $\pm$ 0.06 (syst) $\pm$ 0.22 (br) keV is consistent with the PDG value of 0.47 $\pm$ 0.17 keV \cite{PDG}. 
The PDG average of $\Gamma_{\gamma\gamma}(\chi_{c2})$ neglects correlated systematic uncertainties between experiments.
The NLO PQCD prediction of 0.25 - 0.47 keV is consistent with our measurement and the PDG value. Our measurements show that PQCD based calculations are able to predict the ratios of decay rates, where non-perturbative effects cancel; however the uncertainties in both theory and experiment limit the precision of such comparisons.

We gratefully acknowledge the effort of the CESR staff in providing us 
with excellent luminosity and running conditions. 
This work was supported by the National Science Foundation, 
the U.S. Department of Energy, the Research Corporation, 
the Natural Sciences and Engineering Research Council of Canada 
and the Texas Advanced Research Program.


\begin{thebibliography}{11}
\tighten

\bibitem{Kwong}
W. Kwong {\it et al.}, 
Phys. Rev.  D {\bf 37}, 3210 (1988).

\bibitem{PDG}
Particle Data Group, D. E. Groom {\it et al.}, 
Eur. Phys. J. C {\bf 15}, 1 (2000).

\bibitem{Lee}
R. A. Lee, SLAC-report-282, Stanford University Ph. D. Thesis (1985), 
unpublished.

\bibitem{E835}
M. Ambrogiani {\it et al.}, 
Phys. Rev.  D {\bf 62}, 052002 (2000).

\bibitem{Mangano}
M. L. Mangano and Andrea Petrelli,
Physics Letters {\bf B352}, 445 (1995 ). Inclusion of the color-octet 
contribution was the original idea of E. Braaten.

\bibitem{Huang}
H. W. Huang, C. F. Qiao and K. T. Chao, 
Phys. Rev. D {\bf 54}, 2123 (1996).

\bibitem{CLEOII}
CLEO Collaboration, Y. Kubota {\it et al.}, 
Nucl. Instrum. Methods Phys. Res., Sect. A {\bf 320}, 66 (1992).

\bibitem{CLEOII.V}
T. Hill, 
Nucl. Instrum. Methods Phys. Res., Sect. A {\bf 418}, 32 (1998).

\bibitem{GEANT}
R. Brun {\it et al.}, 
GEANT 3.15, CERN Report No. DD/EE/84-1 (1987).

\bibitem{BGMS}
V. M. Budnev {\it et al.},  
Physics Reports C, {\bf 15}, 181 (1975).

\bibitem{MRK1}
MRKI Collaboration, W. M. Tanenbaum {\it et al.}, 
Phys. Rev. D {\bf 17}, 1731 (1978).

\bibitem{BES}
BES Collaboration, J. Z. Bai {\it et al.}, 
Phys. Rev. D {\bf 60}, 072001 (1999).

\bibitem{XBAL}
Crystal Ball Collaboration, J. E. Gaiser {\it et al.}, 
Phys. Rev. D {\bf 34}, 711 (1986).
\end{thebibliography}
\end{document}